\documentstyle[11pt,cosmos]{article}
\newcommand{\be}{\begin{equation}}
\newcommand{\ee}{\end{equation}}
\newcommand{\ben}{\begin{eqnarray}}
\newcommand{\een}{\end{eqnarray}}
\newcommand{\n}{\label}

\begin{document}

\title{Quintessence Energy and Dissipation}

\author{Luis P. Chimento and Alejandro S. Jakubi}
\affil{Departamento de F\'{\i}sica, Universidad de 
Buenos Aires, 1428 Buenos Aires, Argentina}
\author{Diego Pav\'{o}n}
\affil{Departamento de F\'{\i}sica, Universidad Aut\'onoma 
de Barcelona, 08193 Bellaterra (Barcelona), Spain}

\begin{abstract}
The combined effect of a dissipative fluid and quintessence energy can 
simultaneously drive an accelerated expansion phase at the present time 
and solve the coincidence problem of our current Universe. A solution
compatible with the observed cosmic acceleration is succinctly presented.  
\end{abstract}


\section{Introduction}
In recent years observation of Type Ia supernovae has lend strong 
support to an accelerated expansion of the Universe at present time
[\cite{acc}]. This unexpected feature could be explained by resorting 
to a small cosmological constant $\Lambda$, which on the one hand 
would provide enough negative pressure to account for this acceleration, 
and on the other hand would contribute an energy density of the same order 
of magnitude than the energy density of the matter content 
(baryonic plus dark) of today's Universe -say $\rho_{m} \simeq 0.3$ 
and $\rho_{\Lambda} \simeq 0.7$.  This seemly straightforward ``solution" 
poses, however, a serious problem. Since $\rho_{m}$ redshifts as $a^{-3}$  
while $\Lambda$ is a constant, why they turn out to be comparable today? 
i.e., why we happen to live in a very special (and rather short) 
phase of cosmic expansion? This is the so--called {\it coincidence 
problem} [\cite{stein}].

To avoid it a new form of dark energy (quintessence energy, or Q--matter)
has been introduced. This energy corresponds to a scalar field $\phi$ 
that slowly rolls down its potential with the key property of having a 
negative pressure. In some respects it mimics the scalar field that 
suppossedly drived inflation at the very early Universe -see e.g. 
[\cite{quintessence}]. All these models -formulated for spatially flat
Friedmann--Lema\^{\i}tre--Robertson--Walker (FLRW) universes- overlook 
the fact that the matter component of the Universe (baryonic and 
non--baryonic) might not be very well approximated by a perfect fluid 
since, in general it should behave as a dissipative fluid and 
therefore it must 
have a non--equibrium pressure that might be non--negligible. 

In this short report we shall show that the combination of
quintessence energy (an evolving scalar field with negative pressure) 
and a perfect matter fluid cannot simultaneously drive the current 
accelerated expansion and solve the coincidence problem. However, 
when the matter fluid is no longer assumed perfect this difficulty 
disappears altogether for open and flat FLRW universes [\cite{chimento}].  

At this point it is fair to say, however, that attempts to solve 
the coincidence problem have been also made by using scalar--tensor 
theories of gravity rather than general relativity [\cite{orfeu}]. 
We will not deal with these here.

\section{Quintessence plus perfect fluid}
The stress--energy tensor of a perfect fluid (normal matter, i.e.,
baryonic and non--baryonic) plus a scalar field reads
\be
T_{ab} = (\rho_{m} + \rho_{\phi} + p_{m} + p_{\phi}) u_{a} u_{b} 
+ (p_{m} + p_{\phi}) g_{ab}  \qquad (u^{a} u_{a} = -1).
\label{set}
\ee
The equation of state for the normal matter is 
$p_{m} = (\gamma_{m} - 1) \rho_{m}$, with the baryotropic index lying in
the range $1 < \gamma_{m} < 2$, and 
\be
\label{prhophi}
\rho_{\phi}  =   \frac{1}{2} \dot{\phi}^{2} + V(\phi)\, ,
\qquad
p_{\phi} =   \frac{1}{2} \dot{\phi}^{2} - V(\phi) \, .
\ee
\noindent
A corresponding equation of state for the scalar field can be written
as $p_{\phi} = (\gamma_{\phi} -1) \rho_{\phi}$, or what it is the same

\begin{equation} \label{gammaphi}
\gamma_\phi=\frac{\dot\phi^2}{(\dot\phi^2/2)+V(\phi)},
\end{equation}
\noindent
where for non-negative potentials $V(\phi)$ one has 
$0 \leq \gamma_{\phi} \leq 2$. However, the scalar field can be properly 
interpreted as quintessence if the restriction $\gamma_{\phi} < 1$ is met.

The Einstein equations for any FLRW universe take the form 
\be
\Omega_{m} + \Omega_{\phi}+\Omega_k=1,
\label{constr}
\ee

\be
\dot{\Omega} = \Omega \left(\Omega - 1\right) \left(3 \gamma - 2\right) H \, ,
\label{dOmega}
\ee
\noindent
where $\Omega \equiv \Omega_{m} + \Omega_{\phi}$, 
$\Omega_m \equiv \rho_{m} /\rho_{c}$, 
$\Omega_{\phi} ,\equiv \rho_{\phi} /\rho_{c}$, with
$\rho_{c} \equiv 3 H^{2}$ the critical density, 
and $\Omega_k \equiv-k/(aH)^2$. As usual
$H \equiv \dot{a}/a$ denotes the Hubble factor, and $k$ the spatial 
curvature index.\\
Likewise the evolution equation for the scalar field
$\ddot{\phi}+3H\dot{\phi}+V' = 0$ can be recast as

\be
\dot{\Omega}_{\phi} = \left[2 + \left(3 \gamma - 2\right) \Omega - 3 
\gamma_{\phi}\right] \Omega_{\phi} H \, ,
\label{dOmegaphi}
\ee
\noindent
where $\gamma$ is the average baryotropic index defined by
\begin{equation} 
\gamma\Omega=\gamma_m\Omega_m+\gamma_\phi\Omega_\phi.
\label{gammaOmega}
\end{equation}

The combined measurements of the cosmic microwave background  temperature
fluctuations and the distribution of galaxies on large scales suggest
a flat or nearly flat Universe [\cite{flat}]. Hence the interesting 
solution  of (\ref{dOmega}) at late times is $\Omega = 1$
(i.e., $ k= 0$), and so we discard the solution $\Omega=0$ as incompatible
with observation. The solution $\Omega=1$ will be asymptotically stable
for  expanding universes provided that condition 
$\partial\dot\Omega/\partial\Omega<0$ holds in a neighborhood of $\Omega=1$,
and this implies $\gamma < 2/3$. Hence the matter stress violates the
strong energy condition (SEC) $\rho +3p\ge 0$ and as a consequence
the Universe accelerates its expansion, i.e., 
$3 \ddot{a}/a = -(\rho+3p)/2 > 0$.

Since the mixture of Q--matter and
perfect dark matter fluid must violate the SEC, $\gamma_{\phi}$ must be 
low enough. Namely, because of $\gamma<2/3$, $\gamma_{m}\ge 1$, and
$\gamma_{\phi} < \gamma_{m}$, equation (\ref{gammaOmega}) implies 
$\gamma_{\phi} < \gamma$. Then, introducing $\Omega = 1$ in equation 
(\ref{dOmegaphi}) we obtain

\begin{equation} \label{dOmegaphi2}
\dot{\Omega}_{\phi} = 3 (\gamma - \gamma_{\phi}) \Omega_{\phi} H,
\end{equation}

\noindent 
and therefore $\dot{\Omega}_{\phi} > 0$, i.e., $\Omega_{\phi}$ will
grow until the constraint (\ref{constr}) is saturated, giving 
$\Omega_\phi= 1$ in the asymptotic regime. Thus the matter 
fluid yields a vanishing contribution to the energy density 
of the Universe at large times. This implies that a flat
FLRW universe driven by a mixture of normal perfect fluid and 
quintessence matter cannot both drive an accelerated expansion 
and solve the coincidence problem. Therefore some other 
contribution must enter the stress--energy tensor of the 
cosmic fluid. A sensible choice is a negative pressure 
arising from the dissipative character of the matter component.
It is worth mentioning that in deriving the above result 
neither $\gamma_{m}$ nor $\gamma_{\phi}$ were restricted 
to be constants.

\section{Quintessence plus dissipative fluid}

The only dissipative pressure that may enter the strees--energy 
tensor is a scalar pressure $\pi$ which has to be semi--negative 
definite for expanding fluids to comply with the second law of 
thermodynamics. In this light the stress--energy tensor keeps the
same form as (\ref{set}) but with $p_{m}$ replaced by $p_{m} + \pi$.
Now equation (\ref{constr}) remains in place but 

\be
\label{dOmegapi}
\dot{\Omega} = \Omega \left(\Omega - 1\right)
\left[3 \left(\gamma + \frac{\pi}{\rho}\right) - 2
\right]
H \, ,
\ee

\noindent
and

\be
\label{dOmegaphipi}
\dot{\Omega}_{\phi} = \left\{2 + \left[3 \left(\gamma +
\frac{\pi}{\rho}\right) - 2\right]
\Omega - 3 \gamma_{\phi}\right\} H \Omega_{\phi},
\ee
\noindent
substitute equations (\ref{dOmega}) and (\ref{dOmegaphi}), respectively. 
The energy conservation equation of the normal matter is
$\dot{\rho_{m}} + 3\left(\gamma_{m}+ 
\pi/\rho_{m}\right) \rho_{m} H = 0$.
Owing to the presence of the dissipative pressure $\pi$ the 
constraint $ \gamma < 2/3 $ does not longer have to be fulfilled 
for the solution $\Omega = 1$ of equation (\ref{dOmegapi}) to be
stable. Likewise, inspection of (\ref{dOmegaphipi}) shows that 
when $\Omega = 1$  one can have $\dot{\Omega}_{\phi} <0 $ just 
by choosing the ratio $\pi/\rho$ sufficiently negative. 
Thereby the constraint (\ref{constr}) allows a nonvanishing
$\Omega_{m}$ at large times. By contrast tracker fields based models
(valid only when $\Omega_{k} = 0$) predict that 
$\Omega_{m} \rightarrow 0$ asymptotically. 

A fixed point solution of equation (\ref{dOmegapi}) is $\Omega=1$. Note
that equations (\ref{constr}) and (\ref{dOmegaphipi}) have fixed point 
solutions $\Omega_m=\Omega_{m0}$ and $\Omega_\phi=\Omega_{\phi 0}$,
respectively, when the partial baryotropic indices and the dissipative
pressure are related by

\begin{equation} \label{gammapi}
\gamma_{m}+\frac{\pi}{\rho_{m}}=\gamma_{\phi} = -\frac{2\dot{H}}{3H^2}.
\end{equation}

\noindent
Then, the smaller $\gamma_\phi$, the larger the dissipative effects.
Let us investigate the requirements imposed by the stability of these
solutions. From (\ref{dOmegapi}) we see that $\gamma+\pi/\rho<2/3$ must be
fulfilled if the solution $\Omega=1$ is to be asymptotically stable. This
condition, together with (\ref{gammapi}), leads to the additional constraint
on the viscosity pressure $\pi<\left(2/3-\gamma_m\right)\rho_{m}$,
which must be negative. Also by virtue of (\ref{dOmegaphi2}) and the first 
equality in (\ref{gammapi}) we obtain from last relationship that 
$\gamma_{\phi}<2/3$.

In the special case of a spatially flat universe ($\Omega=1$), the
stability of the solutions $\Omega_{m0}$ and $\Omega_{\phi 0}$ may be studied
directly from (\ref{dOmegaphipi}). Setting 
$\Omega_\phi=\Omega_{\phi0}+\omega$, with $\mid \omega \mid \ll
\Omega_{\phi 0}$, and using (\ref{gammaOmega}) it follows that

\begin{equation} \label{}
\dot\omega=3\Omega_{m} \left(\gamma_m-\gamma_\phi+
\frac{\pi}{\rho_{m}}\right) H \left(\Omega_{\phi 0}+\omega\right).
\end{equation}

\noindent
Accordingly the solution $\Omega=1$, $\Omega_\phi=\Omega_{\phi 0}$ 
is stable for the class of models that satisfies 
$\psi \equiv\gamma_m-\gamma_\phi+(\pi/\rho_m)<0$
and $\psi\to 0$ for $t\to\infty$. Note that this 
coincides with the attractor condition (\ref{gammapi}).

To study the stability of the solutions $\Omega_{m0}$ and
$\Omega_{\phi 0}$ when $k\neq 0$ we introduce the parameter 
$\epsilon \equiv \Omega_{m}/\Omega_{\phi}$.
As it turns out its evolution is governed by 

\begin{equation} 
\dot\epsilon=
-\frac{3H\epsilon}{\Omega_\phi}\left[\frac{2\dot H}{3H^2}+
\gamma_m+\frac{\pi}{\rho_m}+
\left(\frac{2}{3}-\gamma_m-\frac{\pi}{\rho_m}\right)\Omega_k\right],
\label{depsilon2}
\end{equation}

\noindent
and perturbating this expression about the solution 
$\epsilon = \epsilon_{0} \sim {\cal O}(1)$, 
(i.e., using the ansatz $\epsilon=\epsilon_{0}+\delta$ with $|\delta|\ll 1$) 
we obtain with the help of (\ref{gammapi})

\begin{equation}
\n{delta2}
\dot\delta=-\frac{3}{\Omega_\phi}\left(\frac{2}{3}-
\gamma_{\phi}\right)\Omega_{k} H \left(\epsilon_0+\delta\right)
\end{equation}

\noindent
near the attractor. For $\Omega_{k} > 0$ (negatively spatially curved
universes) the ratio $(\Omega_{m}/\Omega_{\phi})_{0}$ is a stable solution. 
For $\Omega_{k} < 0$ one has to go beyond the linear perturbative regime and/or restrict 
the class of models as in the spatially flat case to determine the
stability of the solution. 

We would like to stress that by large times we mean times after the 
cosmological perturbations evolved into the nonlinear regime. Thus 
the structure formation scenario will not be spoiled by the quintessence
field.

Recently there have been claims that CDM should not be a perfect
fluid because it ought to self--interact (with a mean free--path
in the range $1 \, \mbox{kpc} \leq l \leq 1 \, \mbox{ Mpc}$) if one wishes 
to explain the structure of the halos of galaxies [\cite{self}]. It is not 
unreasonable to think that this same interaction lies at the root of the 
dissipative pressure $\pi$ at cosmological scales. Bearing in mind that  
$l = 1/n \sigma $, with $n$ the number density of CDM particles and 
$\sigma$ the interaction cross section, a simple estimation reveals that 
at such scales $l$ is lower than the Hubble distance $H^{-1}$ and 
accordingly  the fluid approximation we are using is valid.

There exist a handful of solutions with the
desired asymptotic properties -see [\cite{chimento}] for details. 
By way of example we just mention 
$a \simeq t^{- 2/\lambda}$ with

\begin{equation} \label{}
\lambda=\frac{1}{2}\left\{-\left(3\gamma + \nu \right)
+\left[\left(3\gamma - \nu \right)^2+
36\gamma_{m} v^2\Omega_m \right]^{1/2}\right\},
\end{equation}

\noindent
where $\nu$ denotes the number of interactions between CDM particles in
a Hubble's time and $v$ the speed of the dissipative signals. 

We may conclude by stressing that both acceleration and coincidence can be
satisfactory explained by a combination of quintessence and dissipative dark
matter. For these models attractor solutions exist with very interesting
properties: an accelerated expansion, spatially flatness and a fixed ratio of
quintessence to dark matter energy density. In consequence, the quintessence
scenario becomes more robust when the dissipative effect of the nonequilibrium
pressure arising in the CDM fluid is allowed into the picture.
 
In [\cite{chimento}] we presented specific models with an ample region in 
the space of out--of--equilibrium thermodynamic parameters satisfying
observational constraints in the asymptotic attractor regime which our
Universe may well be approaching. In a future research we shall aim 
to improve these constraints by analysis of the luminosity 
distance--redshift relation for type Ia supernovae and simulations 
of structure formation that include dissipative effects.

\acknowledgments
This work has ben partially supported by the Spanish Ministry of Education
under grant PB94-0718. LPC and ASJ thank the University of Buenos Aires
for partial support under project TX-93.

\thebibliography
\bibitem{acc}[1]
S. Perlmutter, {\it et al.},
Astrophys. J. 517 (1999) 565 ; A.G. Riess, {\it et al.}, 
{\em ibid} 116 (1998) 1009; P.M. Garnavich, {\it et al.}, 
{\em ibid} 509 (1998) 74.

\bibitem{stein}[2]
P. Steinhardt, in Critical Problems in Physics, V.L. Fitch and 
D.R. Marlow (eds.), Princeton University Press, Princeton, 1997.

\bibitem{quintessence}[3] 
M.S. Turner, and M. White, Phys. Rev. D 56 (1997) R4439;
R.R. Caldwell, R. Dave, and P.J. Steinhardt,
Phys. Rev. Lett. 80 (1998) 1582; I. Zlatev, L. Wang,
and P. Steinhardt, Phys. Rev. Lett. 82 (1999) 896;
P. Steinhardt, L. Wang and I. Zlatev, Phys. Rev. D 59 (1999)
123504; S. Dodelson, M. Kaplinghat and E. Stewart,
[astro--ph/0002360]. 

\bibitem{chimento}[4]
L.P. Chimento, A. Jakubi, and D. Pav\'{o}n,
Phys. Rev. D 62 (2000) 063508. 

\bibitem{orfeu}[5]
S. Sen, and T. Sesadri, [gr-qc/0007079].

\bibitem{flat}[6]
N.A. Bahcall {\it et al.}, Science {\bf 264} (1999) 1481; M. White,
D. Scott, and E. Pierpaoli, [astro-ph/0004385]. 

\bibitem{self}[7]
D.N. Spergel and P.J. Steinhard, Phys. Rev. Lett. 84 (2000) 3760;
J. P. Ostriker, {\it ibid.} 84 (2000) 5258;
S. Hannestad, and R.J. Scherrer, Phys. Rev. D 62 (2000) 043522;
C. Firmani {\em et al.}, [astro-ph/0002376].

\endthebibliography

\end{document}